\documentclass[11pt, oneside]{article}   	

\usepackage{tikz}
\usepackage[bf]{caption}

\pdfoutput=1
\usepackage{jheppub}

\usepackage[english]{babel}
\usepackage[T1]{fontenc}
\usepackage{lmodern}

\usepackage{graphicx}	

\usepackage{amsmath}								
\usepackage{color}
\usepackage{mathtools}
\usepackage{amssymb}

\usepackage{hyperref}

\hypersetup{
    bookmarks=true,%
    colorlinks,%
    citecolor=blue,%
    filecolor=black,%
    linkcolor=blue,%
    urlcolor=blue
}

\usepackage{overpic}
\usepackage{array}
\usepackage{rotating}

\usepackage{amssymb}

\title{A Zoo of Deformed Jackiw-Teitelboim Models near Large Dimensional Black Holes}

                                           \author{Watse Sybesma}
                                           \affiliation{Science Institute,
                                         University of Iceland, Dunhaga 3, 107 Reykjav\'{i}k, Iceland.}

                                           \emailAdd{watse@hi.is}

\abstract{
We consider a charged Lifshitz black hole in the large transverse dimension limit. In this setup, the dynamics near the black hole horizon are shown to be effectively governed by a family of two-dimensional models of dilaton gravity depending on the ratio of the dynamical parameter characterizing the black hole and the dimension of spacetime. 
This family includes the Callan-Giddings-Harvey-Strominger (CGHS) and Jackiw-Teitelboim (JT) models and their charged equivalents.
This family also contains classes of asymptotically anti-de Sitter models beyond JT, characterized by a running Ricci scalar, with the option of adding charge.
Finally, we argue that specific non-minimally coupled probe scalars in the parent Lifshitz model become minimally coupled scalars in the two-dimensional theory, which is relevant for understanding semi-classical corrections in such models.
}
\begin{document}
\maketitle
\newpage
\section{Introduction}
Dilaton models of two-dimensional gravity allow for good analytic control even in the presence of semi-classical corrections. 
However, a priori there are many terms one can write down, which leads to a wealth of possible actions one can use, which can make it hard to justify choosing a specific action, see e.g. \cite{Grumiller:2007ju}.
On the one hand, to constrain the possible actions, it can be convenient to view such models as arising as, e.g., near horizon descriptions of higher dimensional black holes. 
On the other hand, having a higher dimensional picture provides a guideline for in which regime or region of parameter space one should (dis)trust a two-dimensional model and to what higher dimensional quantities certain two-dimensional quantities can be mapped.
In other words, one should be cautious when studying a two-dimensional model of which one is not aware of their higher dimensional pedigree. 
In this note, for the first time, we derive a higher dimensional perspective for a family of two-dimensional theories, that includes the Jackiw-Teitelboim (JT) model \cite{Jackiw:1984je,Teitelboim:1983ux}, by applying the approach of considering a large number of dimensions to so-called Lifshitz black holes.

When studying black holes in $d+2$ dimensions with $d\gg1$, it turns out that the non-trivial, interesting part of gravitational dynamics are strongly localized within a region close by the horizon \cite{Emparan:2013moa}, see \cite{Emparan:2020inr} for a review. Beyond the near horizon region black hole effects are suppressed. Within this near horizon region, \cite{Soda:1993xc,Emparan:2013xia} pointed out the following. For a wide class of black holes, the near horizon dynamics become that of the black hole that arises in the low energy limit of two-dimensional string theory \cite{Elitzur:1990ubs,Mandal:1991tz,Witten:1991yr}, which is governed by the two-dimensional string effective action.

Meanwhile, in the context of Lifshitz holography\footnote{For a review see \cite{Taylor:2015glc}} where one requires the anisotropic scale invariance $t\to \lambda^{z}t$, $\vec{x}\to\lambda\vec{x}$ between time $t$ and space $\vec{x}$, the following was pointed out in several works \cite{Hartnoll:2010gu,Hartnoll:2011dm,Hartnoll:2011fn,Hartnoll:2012rj}. Namely, from the gravitational perspective taking the dynamical critical exponent $z\gg1$ yields physics that is effectively governed by a product space of two-dimensional Anti-de Sitter space ($\text{AdS}_{2}$) and its transverse space. 
However, $z\gg1$ theories are unlikely to describe stable phases of matter down to low temperatures due to their extensive ground state entropy, as can be seen from the following argument. Once we introduce (uncharged) planar black holes in the Lifshitz bulk in order to obtain a temperature, we find that the entropy $S$ is related to temperature $T$ as $S\sim T^{d/z}$, where $d$ is the dimension of the transverse space. 
However, taking $z\gg1$ causes entropy and temperature to cease being related.
In other words, the resulting thermodynamics is limited to share features of an extremal black hole.

In this note we combine $z\gg1$ in the Lifshitz system from a large $d$ perspective, in which we can see different two-dimensional asymptotically $\text{AdS}_{2}$ black holes arising as a near horizon limit, much in the spirit of the aforementioned works \cite{Elitzur:1990ubs,Emparan:2013xia} where the low energy string black hole was obtained. 
To be more specific, we keep $\alpha= z/d$ fixed, we find $S\sim T^{1/\alpha}$ and we show that the following two-dimensional model arises
\begin{equation}\label{eq:model1}\begin{aligned}
	S
	=&~
	\frac{1}{16\pi \tilde{G}_{N}}\int d^{2}x\sqrt{-\gamma}
		e^{-2\phi}\left[
			R_{\gamma}
			+
			4(1-\alpha)(\nabla\phi)^{2}
			+
			4\lambda^{2}
			-
			\frac{1}{4}e^{-4\alpha\phi}F^{2}
		\right]
	\,,
	\quad
	\alpha=\frac{z}{d}
	\,,
\end{aligned}\end{equation}
where $\tilde{G}_{N}$ is a dimensionless parameter, $R_{\gamma}$ the Ricci scalar, $\phi$ a dilaton, $\lambda$ an energy scale and $F$ an electric two-form.
It is emphasized that distinct values of $\alpha$ are not related by Weyl transformations.
We furthermore suppress the boundary terms in this note, as they will not be relevant for our analysis.
Let us turn off the electric field $F$ for now.
It turns out that $\alpha=1$ yields the JT model with a negative cosmological constant, although with the topological term absent, and $\alpha=0$ reproduces the two-dimensional string effective action (which with a slight abuse of nomenclature is also often called the CGHS model as is explained below).\footnote{The topological term usually captures the entropy of the higher dimensional extremal black hole near which the two-dimensional model arises, consider JT arising near a near extremal Reissner-Nordstr\"{o}m black hole for example. The absence of the toplogical term reflects the fact that the large $d$ procedure does not involve a near extremal black hole or at least its entropy is vanishing.} For the values of $\alpha<1$ or $\alpha>1$ we find models with solutions with running Ricci scalars but that are nevertheless asymptotically AdS$_{2}$, see Table \ref{table:1} for an overview. 
\begin{table}[h!]
\centering
\begin{tikzpicture}
\node (table) [inner sep=0pt] {
\begin{tabular}{l|l}
 \textbf{value $\alpha=\frac{z}{d}$} & \textbf{Conformal structure -- specific two-dimensional model}\\
  \hline
  $\alpha=0$ & Schwarzschild, constant Ricci scalar -- string effective action or ``CGHS'' \\
  $0<\alpha<1$ & Schwarzschild-AdS, running Ricci scalar  
  \\
  $\alpha=1$ &  Schwarzschild-AdS, constant Ricci scalar -- JT without topological term\\
  $\alpha>1$ & extremal Reissner-Nordstr\"{o}m, running Ricci scalar   \\
\end{tabular}
};
\draw [rounded corners=.5em] (table.north west) rectangle (table.south east);
\end{tikzpicture}
\caption{Classification of solutions of model presented in \eqref{eq:model1} with the electric field turned off. These were first studied in \cite{lemos94}. When turning on the electric field we get a conformal diagram similar to higher dimensional Reissner-Nordstr\"{o}m with appropriate asymptotics.}
\end{table}\label{table:1}

In absence of the electric field $F$, this two-dimensional model has extensively been studied in \cite{lemos94}, but we present a higher dimensional pedigree which was lacking. Related to that, the model can also be framed as a holographic interpretation in the context of IR descriptions of non-relativistic models. Not only does adding the charge, which also has a higher dimensional pedigree, enrich the structure of the black holes, it also provides the opportunity to study the AdS$_{2}$ throat of these two-dimensional models. 

Callan-Giddings-Harvey-Strominger (CGHS) \cite{Callan:1992rs} added $c$ free massless \textit{minimally} scalar fields to the two-dimensional string effective action such that their conformal anomaly dominates the semi-classical approximation when $c\gg1$. 
Unlike the string effective part of the action, it is unknown how to motivate $c\gg1$ minimally coupled scalars from a higher dimensional perspective. 
We point out that the higher dimensional Lifshitz model allows for adding higher dimensional massless free \textit{non-minimally} coupled scalars that become \textit{minimally} coupled massless free scalars in two dimensions. To be explicit, this means one can motivate adding a term like $-\int d^{2}x\sqrt{-g}\sum_{i=1}^{c}\frac{1}{2}(\nabla f_{i})^{2}$ to the action above in \eqref{eq:model1}, amounting to \eqref{eq:total}.

We conclude this note with some comments on hyperscaling violation and dual theories. The remainder of the note is organized in the following manner. In Section \ref{sec:2} we present the effective picture that arises at large dimensions and derive the two-dimensional effective model in Section \ref{sec:3}. In Section \ref{sec:4} we charge the Lifshitz black hole and subsequently the throat. We conclude in Section \ref{sec:5} with some further comments and an outlook.

\section{The effective picture at large dimensions}\label{sec:2}
In this section we present the geometry and thermodynamics that arises near a Lifshitz black hole for large dimensions.
\subsection{Lifshitz black holes at large $d$ and $z$}
In order to support an asymptotically Lifshitz solution to the Einstein equations it is necessary to consider the presence of matter.
We will postpone introducing the Lagrangian and first consider the metric of a $d+2>3$ dimensional planar Lifshitz black hole \cite{Taylor:2008tg}
\begin{equation}\label{eq:metric1}
	ds^{2}
	=
	-
	\left(\frac{r}{L}\right)^{2z}f(r)dt^{2}
	+
	\frac{L^{2}dr^{2}}{r^{2}f(r)}
	+
	\left(\frac{r}{L}\right)^{2}d\vec{x}_{d}^{2}
	\,,
	\quad
	f(r)
	=
	1
	-
	\left(\frac{r_{0} }{r}\right)^{d+z}
	\,,
\end{equation}
where $L$ is the Lifshitz equivalent of the anti-de Sitter length scale, length scale $r_{0}$ is related to the mass parameter of the theory and determines the location of the horizon. The $d\vec{x}_{d}$ denotes the $d$ dimensional transverse hyper planar space. Furthermore $z$ is the dynamical critical exponent, which can be found to be restricted to $z\geq1$ due to null-energy conditions where $z=1$ reproduces the anti-de Sitter result. We will consider 
\begin{equation}
	\alpha=\frac{z}{d}\,,
\end{equation} 
where we keep $\alpha$ fixed. As a consequence, large $d$ will be matched with large values of $z$.

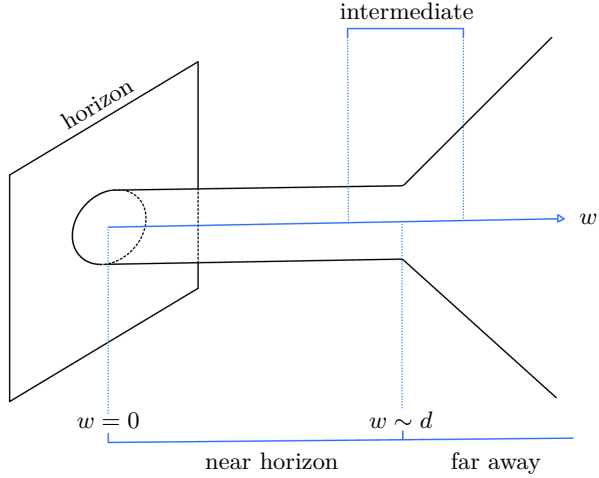
\begin{figure}[h]
	\begin{center}
	\begin{overpic}[width=0.5\textwidth]{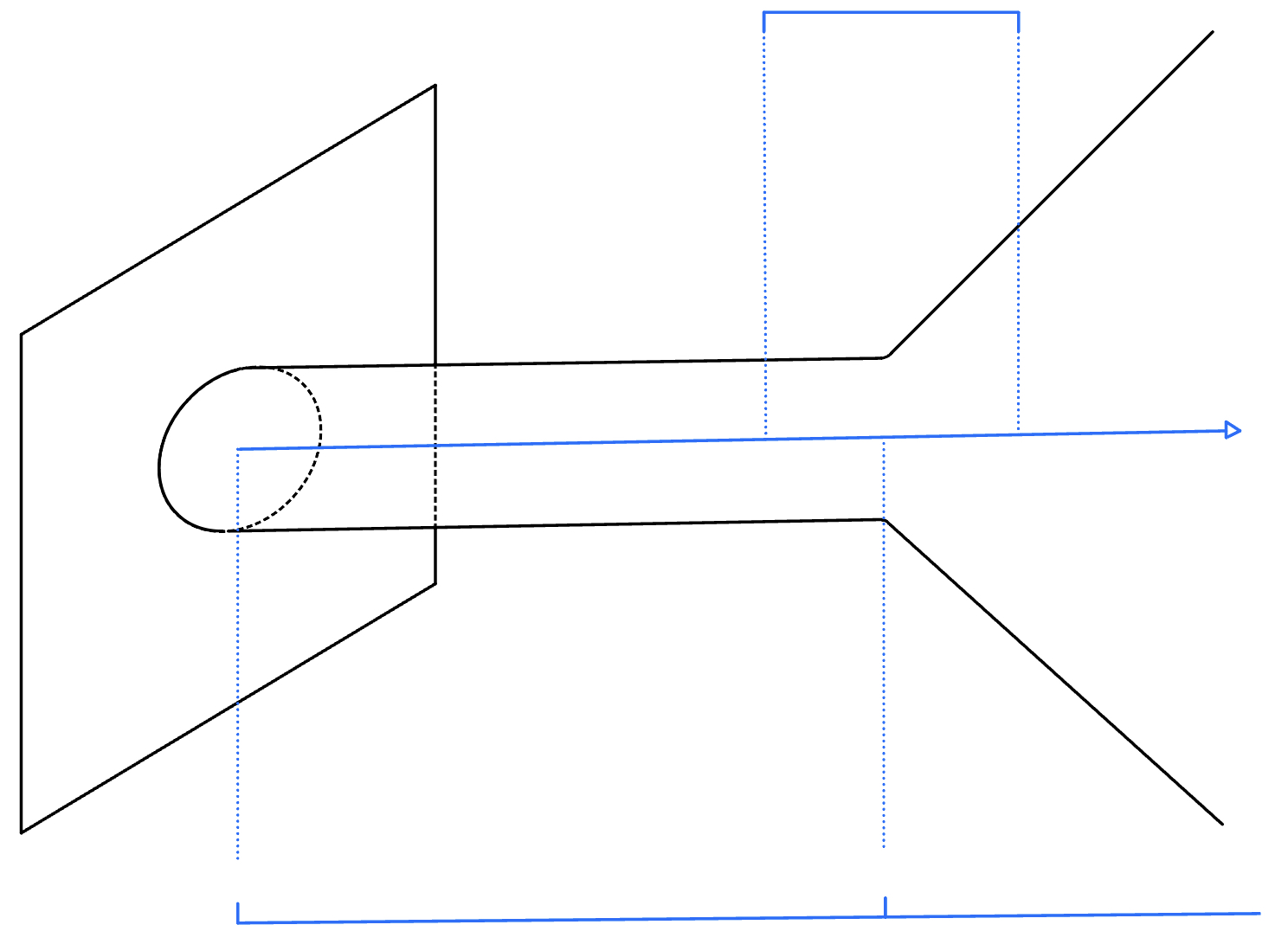}
			\put (13,4) {\rotatebox{0}{\footnotesize{$w=0$}}}
			\put (63,4) {\rotatebox{0}{\footnotesize{$w\sim d$}}}
			\put (10,55) {\rotatebox{31}{\footnotesize{horizon}}}
			\put (58,74) {\rotatebox{0}{\footnotesize{intermediate}}}
			\put (99,39) {\rotatebox{0}{\footnotesize{$w$}}}
			\put (35,-3) {\rotatebox{0}{\footnotesize{near horizon}}}
			\put (77,-3) {\rotatebox{0}{\footnotesize{far away}}}
		\end{overpic}
	\end{center}
\caption{
The near horizon region captures the black hole physics for $d\gg1$. This region corresponds to $r-r_{0}\ll r_{0}$ (or $w\ll d$), whereas the far away region corresponds to $r-r_{0}\gg r_{0}/d$ (or $w\gg1$). In the far away region one finds a Lifshitz vacuum with $z\gg1$. This region is not of interest to us in this note. Notice that there is a range of interpolation, represented by the intermediate region. The relation between coordinates is given in \eqref{eq:xcoord}.}
\label{fig:cartoon1}
\end{figure}

For convenience we introduce a coordinate $w$ that covers the outside of the horizon region:
\begin{equation}\label{eq:xcoord}
	\frac{r}{r_{0}}
	=
	1
	+
	\frac{w}{d}
	\,.
\end{equation}
The real $w$ coordinate allows us to discriminate in which radial region we expect influences of the black hole to remain non-trivial as $d\gg1$.
In combination with $d\gg1$ we consider 
\begin{equation}\label{eq:dubbelucoord}
	w\in[w_{1},w_{2}]
	\quad 
	\text{with} 
	\quad
	0\leq w_{1}<w_{2} 
	\quad\text{and}
	\quad
	w_{1},\,w_{2}\sim d^{n}
	\,.
\end{equation}	
We will be interested in three regions, the far away from the horizon regime $n>1$, the intermediate (or interpolating) regime $n=1$ and the near horizon region $n<1$. It's this last region that will probe black hole physics. See Figure \ref{fig:cartoon1} for a schematic overview.

\paragraph{Far away from the horizon.} For $n>1$ we focus on the far away from the horizon region and obtain\footnote{See Appendix \ref{ap:identity} for details.} 
\begin{equation}
	ds^{2}
	=
	-\nu_{0}^{2\alpha} 
	\left(\frac{w}{d}\right)^{2\alpha d}dt^{2}
	+
	\ell_{d}^{2}\frac{dw^{2}}{\left(\frac{w}{d}\right)^{2}}
	+
	\left(\frac{w}{d}\right)^{2}dx^{2}_{d}
	\,,
\end{equation}
where we introduced the dimensionless number
\begin{equation}\label{eq:nu0}
	\nu_{0}=\left(\frac{r_{0}}{L}\right)^{d}\,,
\end{equation}
and characteristic length scale
\begin{equation}
	\ell_{d}=\frac{L}{d}\,.
\end{equation}
The behavior here is vacuum Lifshitz space with $z\gg1$.

\paragraph{Intermediate.} Taking $n=1$ the intermediate region sits between the far and near region. Its metric reads
\begin{equation}
	ds^{2}
	=
	-\nu_{0}^{2\alpha}
	\left(
		1+\frac{w}{d}
	\right)^{2\alpha d} f(w)dt^{2}
	+
	\ell_{d}^{2}
	\frac{dw^{2}}{\left(1+\frac{w}{d}\right)^{2}}
	+
	\left(1+\frac{w}{d}\right)^{2} d\vec{x}^{2}_{d}
	\,,
\end{equation}
\begin{equation}
	f(w)
	=
	1-\left(1+\frac{w}{d}\right)^{-(1+\alpha)d}
	\,.
\end{equation}
This region interpolates between the vacuum Lifshitz space far away and the near horizon behavior.

\paragraph{Near horizon region.} Near the horizon, for $n<1$ we find that the gravitational dynamics are governed by
\begin{equation}
	ds^{2}
	=
	-\nu_{0}^{2\alpha}
	e^{2\alpha w} f(w)dt^{2}
	+
	\ell_{d}^{2}
	\frac{dw^{2}}{f(w)}
	+
	d\vec{x}^{2}_{d}
	\,,
	\quad
	f(w)
	=
	1-e^{-(1+\alpha)\omega}
	\,.
\end{equation}
For $\omega=0$ we reproduce the coordinate singularity at the horizon. 
We point out that the space becomes a product of a two-dimensional space and the transverse $d$-dimensional hyperplane.
The corresponding Ricci scalar is given by 
\begin{equation}
	R
	=
	\frac{e^{-w(1+\alpha)}(1-\alpha)-2\alpha^{2}}{\ell_{d}^{2}}
	\,,
	\quad
	\alpha
	=
	\frac{z}{d}
	\,.	
\end{equation}
We observe that for the asymptotical value $w\gg1$, only the last term, the curvature of the vacuum, contributes negatively to the Ricci scalar unless $\alpha=0$.
For $\alpha=0$, which corresponds to $z$ finite and only $d\gg1$, this metric is asymptotically flat and in fact the metric is the two-dimensional string black hole as was pointed out in \cite{Emparan:2013xia} for $z=1$.

From the behavior of the Ricci scalar it is suggested that the remaining cases $0<\alpha<1$, $\alpha=1$ and $\alpha>1$ all have characteristically different behavior, despite all being asymptotically Anti-de Sitter. In the case $\alpha=1$, the Ricci scalar does not run and we will in fact show from a reduction that this case is governed by the JT model with a negative cosmological constant. 
In the context of Lifshitz holography it was pointed out before that $\alpha=1$ or $z=d$ enjoys extra symmetries, see e.g. \cite{Sybesma_2015,Gursoy:2016tgf,Keranen:2016ija}.
Note that in all but the $\alpha=1$ case, the Ricci scalar blows up if we would extend $w\to-\infty$, signaling a black hole singularity.

The proper length of the near horizon region, or throat, is not infinite. We compute the proper length $\Delta\ell$ of the throat
\begin{equation}
	\Delta\ell
	=
	\ell_{d}\int^{w_2}_{w_1=0}\frac{dw}{
	\sqrt{1-e^{-(1+\alpha)w}}}
	=
	\ell_{d}\frac{2}{1+\alpha}
	\log\left[
		e^{
		\frac{1+\alpha}{2}w_{2}
		}
		+
		\sqrt{
			e^{(1+\alpha)w_{2}}
			-
			1
		}
	\right]
	\,,
\end{equation}
where $w_{2}$ at its largest is $w_{2}\sim d$. Recalling that $\ell_{d}=L/d$, this shows that $\Delta \ell$ is finite as $d\gg1$. This contrasts the length of the throat of the near horizon near extremal Reissner-Nordstr\"{o}m, which can be extended arbitrarily.

Finally, if we choose $\alpha=0$ the metric approaches Schwarzschild gauge. For $\alpha>0$ we can adopt the coordinate $\rho=\frac{\ell_{d}}{\alpha}\nu_{0}\exp (\alpha \omega)$ to obtain Schwarzschild gauge
\begin{equation}\label{eq:schwarzschild}
	ds^{2}
	=
	-F(\rho)dt^{2}
	+
	\frac{d\rho^{2}}{F(\rho)}
	+
	d\vec{x}^{2}_{d}
	\,,
	\quad
	F(\rho)
	=
	\left(\frac{\alpha}{\ell_{d}}\rho\right)^{2}
	-
	\nu_{0}^{1+\alpha}
	\left(\frac{\alpha}{\ell_{d}}\rho\right)^{\frac{\alpha-1}{\alpha}}
	\,.
\end{equation}
When performing the reduction of the transverse space, we end up with a consistent truncation to a two-dimensional solution that will inherit the $t$ and $\rho$ components presented here.
\subsection{Thermodynamics}
We will establish what thermodynamics we expect in the throat. We take the temperature and entropy of the original planar black hole \cite{Tarrio:2011de} and take the $d\gg1$ limit whilst keeping $\alpha=z/d$ fixed. This results in
\begin{equation}\label{eq:tempent}
	T
	=
	\frac{d+z}{4\pi} \frac{r^{z}_{0}}{L^{z+1}}
	\stackrel{d\gg1}{=}
	\frac{1+\alpha}{4\pi}\nu_{0}^{\alpha}\ell_{d}^{-1}
	\,,
	\quad
	S
	=
	\frac{V_{d}}{4\pi G_{N}} r^{d}_{0}
	\stackrel{d\gg1}{=}
	\frac{\nu_{0}}{4\tilde{G}_{N}}
	\,,	
\end{equation}
where $V_{d}=L^{-d}\int d^{d}x$ is the dimensionless volume factor of the transverse space, $G_{N}$ is the $d+2$ dimensional Newton's constant of the bulk and $\tilde{G}_{N}=G_{N}/(V_{d}L^{d})$ is the dimensionless Newton's constant.
As is suggestive from the Schwarzschild gauge in \eqref{eq:schwarzschild}, $\nu_{0}$ as defined in \eqref{eq:nu0} takes the role of the mass parameter. 
This makes sense as it is related to $r_{0}$, which in term again is related to the mass parameter of the higher dimensional black hole. 
Notice that for $\alpha=0$, temperature $T$ and entropy $S$ cease to be related. 
This is known from the classical black hole studied in the CGHS model, see e.g. \cite{Thorlacius:1994ip}. 
In general, the relation between $T$ and $S$ reflects what was expected in the introduction, i.e.,
\begin{equation}
	T\sim S^{\alpha}
	\,.
\end{equation}
The energy becomes
\begin{equation}\label{eq:energy}
	E
	=
	\frac{V_{d}}{16\pi G_{N}} r^{z+d}_{0}L^{-1-z}d
	\stackrel{d\gg1}{=}
	\frac{\nu_{0}^{1+\alpha}}{4\tilde{G}_{N}}\ell_{d}^{-1}
	\,,
\end{equation}
which (using the first law) gives us the Smarr law 
\begin{equation}
	E=\frac{1}{1+\alpha}ST
	\,,
\end{equation}
and the free energy $F$ reads
\begin{equation}
	F
	=
	-\alpha E
	\,.
\end{equation}
The free energy is always negative and as such the black hole solution dominates, as was already known for the parent theory. 
One can modify the planar topology of the horizon to be spherical. This would amount to adding a term proportional to $r^{-2}$ to the warp factor, of which the effect is flushed out when considering large $d$.
\section{Obtaining a two-dimensional model}\label{sec:3}
In this section we introduce the action that we consider and perform dimensional reduction.
\subsection{Einstein-Maxwell-dilaton model}
The Lifshitz planar black hole \eqref{eq:metric1} is a solution of general relativity with in its matter sector an electric two-form field coupled to a scalar. 
The charge of the two-form field is fixed by requiring the asymptotics of the metric to have Lifshitz behavior and as such is not interpreted as a usual (electric) charge. 
The action of interest, dubbed Einstein-Maxwell-dilaton action, is given by  \cite{Taylor:2008tg}
\begin{equation}\begin{aligned}\label{eq:EMD}
	I
	=
	\frac{1}{16\pi G_{N} }
	\int d^{d+2}x\sqrt{-g}
	&
	\left[
		R
		-
		2\Lambda
		-
		\frac{1}{2}(\nabla \psi)^{2}
		-
		\frac{1}{4}e^{\bar{\lambda}\psi}\bar{F}^{2}
	\right]
	\,,
\end{aligned}\end{equation}
where $\Lambda$ represents the cosmological constant, $\bar{F}$ the field strength, $R$ the Ricci scalar, $g_{\mu\nu}$ the metric tensor and $\psi$ a scalar field. 
In particular
\begin{equation}
	\Lambda
	=
	-\frac{(d+z-1)(d+z)}{2L^{2}}
	\stackrel{d\gg1}{=}
	-
	\frac{(1+\alpha)^{2}}{2\ell^{2}_{d}}
	\,,
	\quad
	\bar{\lambda}
	=
	-
	\sqrt{\frac{2d}{z-1}}
	\stackrel{d\gg1}{=}
	-
	\sqrt{\frac{2}{\alpha}}
	\,,
\end{equation} 
where for future purposes we included the large $d$ expressions and recall that $\alpha=z/d$.
The equations of motion are given by
\begin{equation}
	R_{\mu\nu}
	-
	\frac{2\Lambda}{d}g_{\mu\nu}
	-
	\frac{1}{2}\partial_{\mu}\psi\partial_{\nu}\psi
	-
	\frac{1}{2}e^{\bar{\lambda}\psi}
	\left[
		\bar{F}_{\mu\sigma}\bar{F}_{\nu}{^\sigma}
		-
		\frac{1}{2d}F^{2}g_{\mu\nu}
	\right]
	=
	0
	\,,
\end{equation}
\begin{equation}
	\nabla_{\mu}\left(
		e^{\bar{\lambda}\psi}\bar{F}^{\mu\nu}
	\right)
	=0
	\,,
	\quad
	\Box\psi
	-
	\frac{\bar{\lambda}}{4} e^{\bar{\lambda}\psi}\bar{F}^{2}
	=
	0
	\,.
\end{equation}
Requiring the metric as in \eqref{eq:metric1} we find solutions for the field strength $\bar{F}$ and dilaton $\psi$ to be
\begin{equation}\label{eq:scalar1}
	\bar{F}_{rt}
	=
	\sqrt{\frac{2(z-1)(d+z)}{L^{2}}}
	e^{-\bar{\lambda}\psi}
	\frac{g_{tt}g_{rr}}{\sqrt{-g}} 
	\,,
	\quad
	e^\psi
	=~
	\left(\frac{r}{L}\right)^{\sqrt{2d(z-1)}}
	\,.
\end{equation}
The explicit expression of the scalar field $\psi$ is important as it will fix a relation with the dilaton that arises from the process of dimensional reduction. 
\subsection{Dimensional reduction}
In order to isolate the two-dimensional dynamics we perform a reduction of the transverse to $t$ and $r$ space. 
Explicitly:
\begin{equation}\label{eq:reductionansatz}
	ds^{2}
	=
	\gamma_{ab}dx^{a}dx^{b}
	+
	e^{-\frac{4}{d}\phi(x^{a})}d\vec{x}_{d}^{2}
	\,,
\end{equation}
where $a$, $b$ run over $t$, $r$ and by comparing to Ansatz \eqref{eq:metric1} and scalar solution \eqref{eq:scalar1} we derive
\begin{equation}\label{eq:dilatons}
	e^{-2\phi}
	=
	\left(
		\frac{r}{L}
	\right)^{d}
	=
	e^{\sqrt{\frac{d}{2(z-1)}}\psi}
	\stackrel{d\gg1}{=}
	e^{\frac{\psi}{\sqrt{2\alpha}}}
	\,.
\end{equation}
It is furthermore established that
\begin{equation}\label{eq:fbar}
	\bar{F}_{rt}
	\stackrel{d\gg1}{=}
	\sqrt{2\alpha(\alpha+1)}\ell_{d}^{-1}e^{2\phi+\sqrt{\frac{2}{\alpha}}\psi}\sqrt{-\gamma}
	\,.
\end{equation}
Performing the reduction and taking $d\gg1$ in the Einstein-Maxwell-dilaton action \eqref{eq:EMD} we establish
\begin{equation}\begin{aligned}\label{eq:themodel0}
	I_{\text{2d}}
	=~
	\frac{V_{d}}{16\pi G_{N}}\int d^{2}x\sqrt{-\gamma}e^{-2\phi}
	&
	\left[
		R_{\gamma}
		+
		4
		(\nabla\phi)^{2}
		+
		\ell_{d}^{-2}(1+\alpha)^{2}
	-
	\frac{1}{2} (\nabla\psi)^{2}
	-
	\frac{1}{4} e^{-\sqrt{\frac{2}{\alpha}}\psi}\bar{F}^{2}
	\right]
	\,.
\end{aligned}\end{equation}
To simplify this action we shall use our higher dimensional insight that $\psi$ and $\phi$ are in fact related \eqref{eq:dilatons} and we furthermore integrate out field strength $\bar{F}$ (which is only there to support the geometry), using expression \eqref{eq:fbar}, to obtain:\footnote{for details see Appendix \ref{ap:integrating}}
\begin{equation}\begin{aligned}\label{eq:themodel}
	I_{\text{2d}}
	=&~
	\frac{1}{16\pi \tilde{G}_{N}}\int d^{2}x\sqrt{-\gamma}e^{-2\phi}\left[
		R_{\gamma}
		+
		4
		(1-\alpha)
		(\nabla\phi)^{2}
		+
		4\lambda^{2}
	\right]
	\,,
\end{aligned}\end{equation}
where $\tilde{G}_{N}$ is the dimensionless Newton's constant as introduced below \eqref{eq:tempent} and where we introduced
\begin{equation}
	4\lambda^{2}
	=
	\ell_{d}^{-2}
	(1+\alpha)
	\geq 0
	\,.
\end{equation}
Let us pause and consider the different models, depending on different values of $\alpha=z/d\geq0$. For $\alpha=0$ we find the two-dimensional string effective action. If one were to add $c$ free minimally coupled scalars, it is called the CGHS model. 

In the case of $\alpha=1$, we reproduce the JT model with a negative cosmological constant, but without the usual topological term that arises when one derives the JT model from considers, e.g., radially infalling modes near the horizon of a near extremal Reissner-Nordstr\"{o}m black hole in four dimensions. The JT model without topological term, is also derived when one considers, e.g., a circular reduction of a BTZ black hole for negative cosmological constant \cite{Achucarro:1993fd} or a circular reduction of three-dimensional de Sitter for a positive cosmological constant \cite{Sybesma:2020fxg}. 

In the case where $\alpha\neq0,1$, we find asymptotically anti-de Sitter solutions, which differ from the usual JT solutions due to a running Ricci scalar. In Table \ref{table:1} in the introduction, we present the characteristic solutions to this model as obtained in \cite{lemos94}. 
The explicit solutions to the model we obtained in \eqref{eq:themodel}, but, we stress, not a higher dimensional pedigree, are given by \cite{lemos94} 
\begin{equation}
	e^{-2\phi}
	=
	(a\rho)^{\frac{1}{\alpha}}
	\,,
	\quad
	ds^{2}
	=
	-\left[
		a^{2}\rho^{2}
		-
		b(a\rho)^{\frac{\alpha-1}{\alpha}}
	\right]dt^{2}
	+
	\frac{d\rho^{2}}{a^{2}\rho^{2}
		-
		b(a\rho)^{\frac{\alpha-1}{\alpha}}}
		\,,
\end{equation}
\begin{equation}
	R_{\gamma}
	=
	-2\lambda^{2}
	\left[
		\frac{4\alpha^{2}}{\alpha+1}
		+
		2b\frac{\alpha-1}{\alpha+1}(a\rho)^{-\frac{\alpha+1}{\alpha}}
	\right]
	\,,
\end{equation}
where $b$ is proportional to the ADM mass and where 
\begin{equation}
	a=\frac{2\alpha\lambda}{\sqrt{(\alpha+1)}}
	=
	\frac{\alpha}{\ell_{d}}
	\,.
\end{equation} 
Note that solving $a\rho=b^{\alpha/(\alpha+1)}$ gives the radius of the horizon. Comparing with our higher dimensional Schwarzschild gauge result \eqref{eq:schwarzschild} and the fact that we are considering a consistent truncation, we establish that
\begin{equation}
	b
	=
	\nu_{0}^{1+\alpha}
	\,.
\end{equation} 
Note that $\alpha=0$ forms a seperate case that was already analyzed in \cite{Emparan:2013xia} for $z=1$.

Furthermore, for $\rho\to0$ one detects the singularity, as is signaled by $e^{-2\phi}\to0$ or the Ricci scalar blowing up. For $\rho\to\infty$ we find $e^{-2\phi}\to\infty$ and the Ricci scalar becoming a negative constant, implying a conformal boundary. 
Let us finally check the resulting entropy. The gravitational entropy becomes
\begin{equation}
	S
	=
	\left.\frac{1}{4\tilde{G}_{N}}e^{-2\phi}\right|_{\text{horizon}}
	=
	\frac{\nu_{0}}{4\tilde{G}_{N}}
	\,,
\end{equation}	
which reproduces the entropy given in \eqref{eq:tempent} as expected.
%
\section{Adding charge to the mix}\label{sec:4}
It is possible to add charge to the Lifshitz black hole, which will be inherited by the two-dimensional picture. Let us first revisit the higher dimensional picture before reducing to two dimensions.
\subsection{Charging the black hole}
Let us start by considering the geometry. Adding charge will only modify the warp factor of the metric we introduced in \eqref{eq:metric1} \cite{Tarrio:2011de}
\begin{equation}
	f(r)
	=
	1
	-
	\left(\frac{r_{0} }{r}\right)^{d+z}
	+
	\frac{q^{2}}{r^{2(d+z-1)}}
	\,,
\end{equation}
where $q$ is the charge parameter of the black hole. We now introduce $r_{h}$, which is the black hole horizon that satisfies $f(r_{h})=0$ and in general $r_{h}\neq r_{0}$. We will now show that we can nevertheless keep on using the $w$ coordinate as introduced in \eqref{eq:xcoord}, $(r/r_{0})^{d}=1+w/d$ to analyze the throat region.
To do so, we establish that
\begin{equation}
	f(r_{h})=0
	\,,
	\quad
	\Rightarrow
	\quad
	\frac{r_{0}^{d+z}}{r_{h}^{d+z}}\stackrel{d\gg1}{=}\xi^{1+\alpha} :=
	\frac{1\pm\sqrt{1-4Q^{2}\nu_{0}^{-2(1+\alpha)}}}{2Q^{2}\nu_{0}^{-2(1+\alpha)}}
	\,,
\end{equation}
where $Q=q/L^{d+z-1}$ is the dimensionless charge and we choose the lower sign for $\xi$ as it corresponds to the outer horizon. We furthermore see that $4Q^{2}\nu_{0}^{-2(1+\alpha)}=1$ implies extremality, as both horizons overlap. Furthermore this implies $\xi^{1+\alpha}=2$ at extremality. 
It can be checked that this satisfies $f'(r_{h})=0$.
The above teaches us that using the coordinate $w$, only $(r/r_{0})^{d}$ will receive a correction of factor $\xi$, but otherwise we can use $r_{0}$ and $r_{h}$ interchangeably.
Let us now turn to the expressions for temperature $T$ and entropy $S$, of which we use the expressions from \cite{Tarrio:2011de}.
The entropy and temperature now become
\begin{equation}
	S
	=
	\frac{V_{d}}{4\pi G_{N}} r^{d}_{h}
	\stackrel{d\gg1}{=}
	\frac{\nu_{0}}{4\tilde{G}_{N}}
	\frac{1}{\xi}
	\,,
\end{equation}
\begin{equation}\begin{aligned}
	T
	=&~
	\frac{r_{h}^{z}}{4\pi L^{1+z}}
	\left[
		(d+z)
		-
		(d+z-2)
		\frac{q^{2}}{r_{h}^{2(d+z-1)}}
	\right]
	\\
	\stackrel{d\gg1}{=}&~
	\xi^{-\alpha} \frac{\nu_{0}^{\alpha}}{4\pi}\ell^{-1}_{d}
	\left[
		1
		-
		\xi^{2(1+\alpha)}\nu_{0}^{-2(1+\alpha)}Q^{2}
	\right]
	\,,
\end{aligned}\end{equation}
which shows that at extremality, $4Q^{2}\nu_{0}^{-2(1+\alpha)}=1$ and $\xi^{1+\alpha}=2$, the temperature indeed vanishes while the entropy remains finite. 
We now check stability in the free energy $F$ in the canonical ensemble (fixed charge) and Gibbs free energy $W$ in the grand canonical ensemble (fixed electric potential):
\begin{equation}
	F
	\stackrel{d\gg1}{=}
	-\frac{\xi^{-1}\nu_{0}}{4\pi \tilde{G}}
	\left[
		\frac{\xi^{-\alpha}\nu_{0}^{\alpha}}{4\pi}\ell_{d}^{-1}
		+
		T
	\right]
	\,,
	\quad
	W
	\stackrel{d\gg1}{=}
	-\frac{\alpha}{4\tilde{G}}\xi^{-1}\nu_{0}T
	\,,
\end{equation}
which are, as expected from higher dimensional considerations, both negative for all attainable temperatures and where we used expressions obtained from \cite{Tarrio:2011de}. We do not expect that taking into account spherical topology will cause phase transitions, as those only arise for $z<2$ and now $z\gg1$.

To conclude this subsection, using $\rho=\frac{\ell_{d}}{\alpha}\nu_{0}\exp (\alpha \omega)$ in the throat region to obtain Schwarzschild gauge as before in \eqref{eq:schwarzschild} we now find
\begin{equation}\label{eq:rnbh}
	ds^{2}
	=
	-F(\rho)dt^{2}
	+
	\frac{d\rho^{2}}{F(\rho)}
	+
	d\vec{x}^{2}_{d}
	\,,
	\quad
	F(\rho)
	=
	\left(\frac{\alpha}{\ell_{d}}\rho\right)^{2}
	-
	\nu_{0}^{1+\alpha}
	\left(\frac{\alpha}{\ell_{d}}\rho\right)^{\frac{\alpha-1}{\alpha}}
	+
	Q^{2}
	\left(\frac{\alpha}{\ell_{d}}\rho\right)^{-\frac{2}{\alpha}}
	\,.
\end{equation}
Next we consider the dimensional reduction.
\subsection{Charging the throat}
The way we charge the Lifshitz black hole is by adding a two-form $F$ that is coupled to scalar $\psi$ in the following specific way \cite{Tarrio:2011de}
\begin{equation}\begin{aligned}\label{eq:twotwoforms}
	I
	=
	\frac{1}{16\pi G_{N} }
	\int d^{d+2}x\sqrt{-g}
	&
	\left[
		R
		-
		2\Lambda
		-
		\frac{1}{2}(\nabla \psi)^{2}
		-
		\frac{1}{4}e^{\bar{\lambda}\psi}\bar{F}^{2}
		-
		\frac{1}{4}e^{\lambda_{Q}\psi}F^{2}
	\right]
	\,,
\end{aligned}\end{equation}
where exponent
\begin{equation}
	\lambda_{Q}
	=
	\sqrt{\frac{2(z-1)}{d}}
	\stackrel{d\gg1}{=}
	\sqrt{2\alpha}
	\,,
\end{equation}
is not equal to $\bar{\lambda}$.
We emphasize that the action in \eqref{eq:twotwoforms} contains two two-forms, of which $\bar{F}$ is only there to support the Lifshitz geometry and $F$ provides the electric charge. The two-forms can be told apart by their different manners of coupling to the scalar $\psi$.
Solving the equations of motion one finds
\begin{equation}\label{eq:electric}
	F_{rt}
	=
	qL^{-d-z+1}\sqrt{
	\frac{2d(d+z-2)}{L^{2}}}
	e^{-\lambda_{Q}\psi}
	\frac{g_{tt}g_{rr}}{\sqrt{-g}} 
		\stackrel{d\gg1}{=}
	Q\sqrt{2(\alpha+1)}\ell_{d}^{-1}e^{2\phi-\sqrt{2\alpha}\psi}\sqrt{-\gamma}
	\,.
\end{equation}
Taking the same reduction as in \eqref{eq:reductionansatz}, $d\gg1$ and the relation between the dilatons \eqref{eq:dilatons}, as in the uncharged case, we finally obtain
\begin{equation}\label{eq:chargedsa}
	I
	=~
	\frac{1}{16\pi \tilde{G}_{N}}\int d^{2}x\sqrt{-\gamma}e^{-2\phi}\left[
		R_{\gamma}
		+
		4
		(1-\alpha)
		(\nabla\phi)^{2}
		+
		4\lambda^{2}
		-
		\frac{1}{4}
		e^{-4\alpha\phi}F^{2}
	\right]
	\,.
\end{equation}
One could choose to integrate out $F$ using \eqref{eq:electric}.
Solving the equations of motion in Schwarzschild gauge and recycling the value for the dilaton we established before, we find 
\begin{equation}
	e^{-2\phi}
	=
	(a\rho)^{\frac{1}{\alpha}}
	\,,
	\quad
	ds^{2}
	=
	-f(\rho)dt^{2}
	+
	\frac{d\rho^{2}}{f(\rho)}
		\,,
		\quad
		f(\rho)
		=
		a^{2}\rho^{2}
		-
		\nu_{0}^{1+\alpha}(a\rho)^{\frac{\alpha-1}{\alpha}}
		+
		Q^{2}\rho^{-\frac{2}{\alpha}}
		\,,
\end{equation}
with $a=
	\alpha/\ell_{d}$. 
Contrasting the uncharged case, this model was not considered by \cite{lemos94}. The global causal structure is (AdS-)Reissner-Nordstr\"{o}m.
In fact, this model allows one to study the charged throat of the large $d$ Lifshitz black hole. Curiously, one could study the extremal limit, so another throat, within the already existing throat solution -- a true nod to \textit{inception}. 
\section{Comments and outlook}\label{sec:5}
We studied a charged Lifshitz black hole model for large $d$ while keeping $\alpha=z/d$ fixed. In this setup, the dynamics near the black hole horizon turn out to be governed by a family of models \eqref{eq:chargedsa}, depending on $\alpha$, which are all asymptotically AdS$_{2}$ (apart from $\alpha=0$ that was obtained before in this manner in \cite{Emparan:2013xia} for $z=1$) and include the JT model with $\alpha=1$ as a special case. The uncharged case was studied before -- but no higher dimensional derivation was given -- in \cite{lemos94}.
For $\alpha\neq0,1$ even the uncharged models have a running Ricci scalar. Below we give some comments on these results.

\paragraph{Minimally coupled scalars.} CGHS added $c$ free massless \textit{minimally} scalar fields to the two-dimensional string effective action such that their conformal anomaly dominates the semi-classical approximation when $c\gg1$. 
Unlike the string part of the action, for $c\gg1$ it was unknown how to motivate these minimally coupled scalars from a higher dimensional perspective. 
Owing to the fact that we are considering a consistent truncation, we know from \eqref{eq:dilatons} that $e^{-2\phi}=\exp[\sqrt{\frac{d}{2(z-1)}}\psi]$. As a result, adding $c$ massless probe scalars $f_{i}$ that do not depend on the transverse space and are of the following specific form yields the minimally coupled two-dimensional scalars:
\begin{equation}\begin{aligned}
	I_{\text{probe}}
	=&
	-\frac{1}{16\pi G_{N}}\int d^{d+2}x\sqrt{-g}e^{-\sqrt{\frac{d}{2(z-1)}}\psi}
	\sum_{i=1}^{c}
	\frac{1}{2}\partial_{\mu} f_{i}\partial^{\mu}f_{i}
	\\
	\stackrel{d\gg1}{=}&
	-\frac{1}{16\pi \tilde{G}_{N}}\int d^{2}x\sqrt{-g}
	\sum_{i=1}^{c}
	\frac{1}{2}\partial_{\mu} f_{i}\partial^{\mu}f_{i}
	\,.
\end{aligned}\end{equation}
Combining this with \eqref{eq:chargedsa} explicitly gives
\begin{equation}\label{eq:total}
	I
	=
	\int d^{2}x\sqrt{-\gamma}
	\left(
		e^{-2\phi}\left[
		R_{\gamma}
		+
		4
		(1-\alpha)
		(\nabla\phi)^{2}
		+
		4\lambda^{2}
		-
		\frac{1}{4}
		e^{-4\alpha\phi}F^{2}
	\right]
	-
	\frac{1}{2}\sum_{i=1}^{c}
	(\nabla f_{i})^{2}
	\right)
	\,.
\end{equation}
These probe scalars on the higher dimensional Lifshitz black hole, from a perspective of model building, generate the CGHS procedure for the broad family of charged and uncharged black holes parametrized by $\alpha=z/d$. A two-dimensional topological term could, perhaps, also be engineered in such fashion.

\paragraph{Hyperscaling violation.} One can further deform the Lifshitz metric to also include an overall term $r^{-2\theta/d}$ where $\theta<d$. This modifies the warp factor to be $f(r)=1+k \frac{(d-1)^{2}}{(d-\theta+z-2)^{2}}-(r_{0}/r)^{d-\theta+z}+q^{2}/r^{2(d-\theta+z-1)}$ \cite{Charmousis:2010zz,Alishahiha:2012qu,Pedraza:2018eey}, where $k=0,1$ corresponds to planar and spherical horizon topology respectively. 
If one keeps $\theta$ of order one, it turns out that this is equivalent to $\theta=0$ in the large $d$ approach.
Requiring, instead $\theta\sim d$, the overall term in the metric $r^{-2\theta/d}$ does not contribute to the near horizon geometry but does modify the resulting two-dimensional warp factor. However, it is expected that the functional form of the resulting two-dimensional model is the same. In \cite{Hartnoll:2012wm} an approach with large $\theta$ was considered.

\paragraph{Quasinormal modes and AdS/CMT.}  The two-dimensional model can also be viewed as a holographic interpretation in the context of IR descriptions of non-relativistic condensed matter models. The large dimension approach has successfully been used to analyze quasinormal modes \cite{Emparan:2014aba,Emparan:2015gva}, something that could complement the numerical results of \cite{Sybesma_2015} and connect to (poles of) two-point functions \cite{Kachru:2020gmz}.

\paragraph{Holographic dual.} The JT model allows for a boundary description through the Schwarzian \cite{Engelsoy:2016xyb}. Meanwhile, it has also been shown that the dual theory allows for the description as an ensemble average \cite{Saad:2019lba}, something that is under investigation for higher dimensional theories. The models with $\alpha\neq0,1$ are asymptotically AdS$_{2}$, but without a vanishing bulk term on-shell, implying an extra contribution for the boundary Schwarzian.\footnote{Despite not stated explicitly in this note, the Gibbons-Hawking-York term takes the same functional as for JT or CGHS.} It would be curious to find out the implication for the matrix model approach of describing this family of models.

\section*{Acknowledgements}
It is a true privilege and an absolute luxury to have had Roberto Emparan, Daniel Gr\"{u}miller, Diego Hidalgo, Thomas Mertens, Juan Pedraza and Andy Svesko take time from their busy schedules to provide feedback on the draft in various stages of the project. I am grateful, thank you. I also want to acknowledge constructive discussions that took place in the group in Iceland.
I am supported by the Icelandic Research Fund via the Grant of Excellence titled ``Quantum Fields and Quantum Geometry''.\\
It brings inexplicable joy to be able to thank you, dear Katla, for messing up my planning.

\newpage
\appendix
\section{Coordinates and their large $d$ behavior}\label{ap:identity}
Let $y$ be a real number that is independent of $d$.
We make use of the identity 
\begin{equation}
	\left(
		1
		+
		d^{n}\frac{y}{d}
	\right)^{-d}
	\stackrel{d\gg1}{=}
	\begin{cases}
		e^{-d^{n}y}\,,\hspace{1.9cm}\text{if }n<1\,,\\
		(1+y)^{-d}\,,\hspace{1.3cm}\text{if }n=1\,,\\
		\left(d^{n-1}y\right)^{-d}\,,\hspace{1.1cm}\text{if }n>1\,,
\end{cases}
\end{equation}
where $y$ is a real parameter that is independent of $d$. As a result, taking $w$ as defined in \eqref{eq:xcoord} and \eqref{eq:dubbelucoord}, we find
\begin{equation}
	\left(
		\frac{r}{r_{0}}
	\right)^{-d}
	=
	\left(
		1
		+
		\frac{w}{d}
	\right)^{-d}
	\stackrel{d\gg1}{=}
	\begin{cases}
		e^{-w}\,,\hspace{2.1cm}\text{if }n<1\,,\\
		\left(1+\frac{w}{d}\right)^{-d}\,,\hspace{1cm}\text{if }n=1\,,\\
		\left(\frac{w}{d}\right)^{-d}\,,\hspace{1.7cm}\text{if }n>1\,,
\end{cases}
\end{equation}
where the $n=1$ case interpolates between $n<1$ and $n>1$.
\section{Integrating out $\bar{F}$}\label{ap:integrating}
In order to go from the two-dimensional model in \eqref{eq:themodel0} to the two-dimensional model in \eqref{eq:themodel}, we need to integrate out the field $\bar{F}$ in the action.We will use the solution of $\bar{F}$ given in \eqref{eq:fbar}.
\begin{equation}
	\int d^{2}x\sqrt{-\gamma}
	\left[
		-\frac{1}{4}e^{-2\phi-\sqrt{\frac{2}{\alpha}}}\bar{F}^{2}
	\right]
	\to
	-
	\int d^{2}x\sqrt{-\gamma}
	\left[
		-\frac{1}{4}e^{-2\phi-\sqrt{\frac{2}{\alpha}}}(-\gamma)(2g^{rr}g^{tt})\bar{F}_{rt}^{2}
	\right]
	=(*)
	\,,
\end{equation}
where the overall minus sign arises because we are integrating out a kinetic term, in order to ensure that the equations of motions stay the same. We continue the computation by inserting the on-shell solution of the field strength
\begin{equation}
	(*)
	=
	\int d^{2}x\sqrt{-\gamma}
	\left[
		-\alpha(\alpha+1)\ell^{-1}_{d}e^{2\phi+\sqrt{\frac{2}{\alpha}}}
	\right]
	=
	\int d^{2}x\sqrt{-\gamma}e^{-2\phi}
	\left[
		-\alpha(\alpha+1)\ell^{-1}_{d}
	\right]
	\,,
\end{equation}
where in the last line we used the relation between $\psi$ and $\phi$ from \eqref{eq:dilatons}. 
Combining this term with the contribution coming from the cosmological constant we obtain $\lambda^{2}$ as given in \eqref{eq:themodel}.

\newpage
\bibliographystyle{JHEP}
\bibliography{refds}
\end{document}